# Intelligent reflecting surface aided wireless networks: Harris Hawks optimization for beamforming design

Huaqiang Xu*
School of Physics and Electronics
Shandong Normal University
Jinan, China
E-mail: xuhq@sdnu.edu.cn

Guodong Zhang
School of Information Science and Technology
Nantong University
Nantong, China
E-mail: gdzhang@ntu.edu.cn

Jun Zhao*
School of Computer Science and Engineering
Nanyang Technological University
Singapore
E-mail: JunZhao@ntu.edu.sg

Quoc-Viet Pham
Research Institute of Computers, Information and Communication
Pusan National University
South Korea
E-mail: vietpq@pusan.ac.kr

*Abstract*—Intelligent Reflecting Surface (IRS) is envisioned to be a promising green and cost-effective solution to enhance wireless network performance by smartly reconfiguring the signal propagation. In this paper, we study an IRS-aided multiple-input single-output wireless network where a multi-antenna Access Point (AP) services a single-antenna user assisted by an IRS. The goal is to maximize the received signal power by jointly optimizing the transmit beamforming at the AP and the reflection coefficient at the IRS. The formulated optimization problem is non-convex and subject to constraints. We adopt a novel nature-inspired optimization technique named Harris Hawks Optimizer (HHO) to tackle the problem. After transforming the constrained problem into an unconstrained problem using penalty method, the formulated problem is optimized by the HHO. To the best of our knowledge, this is the first time to use a meta-heuristic algorithm to solve the IRS-aided network optimization problem. Simulation is conducted to verify the feasibility of the HHO-based scheme. The results show that the HHO-based scheme could provide similar or even better optimization results compared with other optimization algorithms.

## I. INTRODUCTION

The spectrum efficiency of wireless networks has been improved significantly in the last few decades thanks to various novel technologies such as massive Multiple-Input Multiple-Output (MIMO), millimeter wave communication and polar code. This also promotes the improvement in network capacity and data rate, which is in line with the vision of the forthcoming fifth-generation (5G) and beyond wireless networks. However, network energy consumption and hardware cost are still critical issues for realizing sustainable and green 5G networks [1, 2].

Recently, Intelligent Reflecting Surface (IRS) has been proposed as a promising solution to achieve both energy and spectral efficiency with low hardware cost [3]. Specifically, IRS is a reflect array composed of a large number of low-cost, passive unite, where each one is able to independently reflect the incident wireless signal with certain phase shift in a software-defined manner and all of them can collaboratively alter the reflected signal propagation to achieve desired channel response. Different from conventional active mechanisms such as amplify-and-forward relay, IRS passively reflects signals instead of signal-regeneration, thus almost without consuming any energy or requiring additional time/frequency resources for signal propagation. Furthermore, IRS possesses other advantages (e.g., such as low profile, lightweight and easy-deployment), which enables IRS has the potential to be wildly deployed in various application scenarios, such as IRS-aided wireless network [4], IRS-aided wireless power transfer [5] and IRS-aided UAV communication network [6].

In the IRS-aided wireless system, the user could receive signals from both direct (AP-user) link and indirect (IRS-user) link. By adaptively adjusting the phase shifts of all reflecting elements, the reflected signals can be combined coherently at the intended receiver to improve the received signal power or destructively at the non-intended receiver to enhance security. Thus, the optimization of the IRS-aided wireless system is always preferred to achieve a higher performance gain. It is typically formulated into a joint optimization problem of the AP's active transmit beamforming and the IRS's passive beamforming. However, the formulated optimization problem is shown to be non-convex and thus difficult to be solved optimally [7]. The main solution methods in current literature are mostly based on alternating optimization and semi-definite relaxation method, which can guarantee the convergence to sub-optimal solutions [4, 7-9]. The research on performance optimization of the IRS-aided system is still in its infancy, and more optimization techniques need to be developed in the future to achieve a higher performance gain.

This paper adopts a meta-heuristic, namely Harris Hawks Optimizer (HHO) [10] to solve the non-convex optimization problem formulated from the IRS-aided wireless network. The HHO is a novel population-based optimizer inspired by the cooperative behaviors of Harris hawks, an intelligent bird, in hunting the escaping prey. In this paper, we consider a typical IRS-aided MISO wireless network and formulate an optimization problem to maximize the received signal power by jointly optimizing the transmit beamforming vector at the AP and the reflection coefficient vector at the IRS. We adopt the HHO algorithm to tackle the formulated non-convex optimization problem. Compared with other conventional algorithms [4, 7-9], the core operation of the HHO does not rely on objective mathematical traits, thus it is simple and easy in implementation. It is shown by numerical results that the HHO converges to the optimal solution and achieves similar or better optimization results compared with conventional algorithms. To the best of our knowledge, this is the first time to use a meta-heuristic optimizer to solve joint beamforming design problem in the IRS-aided wireless network.

## II. SYSTEM MODEL AND PROBLEM FORMULATION

This paper considers an IRS-aided MISO wireless network [8]. An AP equipped with $M$ antennas provides communication service for a single-antenna user, and the service is assisted by an IRS composed of $N$ passive reflecting elements. Controlled by a smart controller, IRS can work in two modes, the receiving mode for channel sensing and reflecting mode for scattering the incident signals with certain phase shift on them based on the propagation environment learned by the receiving model. All channels involved in this wireless system are assumed to follow a quasi-static flat-fading channel model. Consider severe path loss, the signals reflected by the IRS two or more times are neglected. To focus on the performance enhancement of the IRS-aided wireless network, we assume that the global channel state information is perfectly known at the AP/IRS for joint active and passive beamforming design on them.

The channel coefficients of the AP-user link, AP-IRS link, and IRS-user link are denoted by $\boldsymbol{h}_d^H \in \mathbb{C}^{1 \times M}$, $\boldsymbol{G} \in \mathbb{C}^{N \times M}$ and $\boldsymbol{h}_r^H \in \mathbb{C}^{1 \times N}$, respectively. Here, the superscript $H$ represents the conjugate transpose operation, and $\mathbb{C}^{a \times b}$ denotes the set of $a \times b$ complex valued matrices. The AP transmits an independent and random message $s$ with zero mean and unit variance via linear beamforming. The transmit beamforming vector is denoted by $\boldsymbol{w} \in \mathbb{C}^{M \times 1}$, which is constrained by

$$\| \boldsymbol{w} \|^2 \leq P_{AP}, \tag{1}$$

where $\|\cdot\|$ denotes the Euclidean norm of a complex vector, and $P_{AP}$ is the maximum transmit power at the AP. At the IRS, each reflecting element combines the incident signals from multi-path and re-scatters the combined signal with adjustable phase shift. For the sake of maximizing signal reflecting, the amplitude of the reflection is fixed at 1. Let $\boldsymbol{\theta} = [\theta_1, \cdots, \theta_N]$ denotes the phase shift vector on the combined signal, where $\theta_n \in [0, 2\pi]$. Furthermore, we use $\Theta = \text{diag}(e^{j\theta_1}, \cdots, e^{j\theta_n}, \cdots, e^{j\theta_N})$ to represent the diagonal phase shift matrix, where $j$ denotes the imaginary unit and diag() denotes a diagonal matrix whose diagonal element represents the phase shift of the corresponding reflecting element.

In the IRS-aided wireless network, the user receives signals from both the AP-user link and the IRS-user link. Specifically, the IRS-use link is dominated by the AP-IRS link and the IRS reflecting with phase shifts. Therefore, the channel between the AP and the user can be modeled as a concatenation of the above three components. With given sending signal $s$ at the AP, the total received signal $y$ at the user can be expressed as

$$y = (\boldsymbol{h}_r^H \Theta \boldsymbol{G} + \boldsymbol{h}_d^H) \boldsymbol{w} s + z, \tag{2}$$

where $z$ denotes the Gaussian noise at the user with mean zero and variance $\sigma^2$. Accordingly, the signal power received at the user is

$$\gamma = |(\boldsymbol{h}_r^H \Theta \boldsymbol{G} + \boldsymbol{h}_d^H) \boldsymbol{w}|^2. \tag{3}$$

In practice, information transmission benefits from stronger signal power. Thus, our objective is to maximize the received signal power by jointly optimizing the transmission beamforming vector $\boldsymbol{w}$ at the AP and the phase shift vector $\boldsymbol{\theta}$ at the IRS, subject to the maximum transmit power constraint at the AP. The optimization problem is formulated as

$$\begin{aligned} \max_{\boldsymbol{w}, \boldsymbol{\theta}} \quad & |(\boldsymbol{h}_r^H \Theta \boldsymbol{G} + \boldsymbol{h}_d^H) \boldsymbol{w}|^2 \\ \text{s.t.} \quad & \| \boldsymbol{w} \|^2 \leq P_{AP}, \\ & 0 \leq \theta_n \leq 2\pi, \forall n = 1, \cdots N. \end{aligned} \tag{4}$$

The formulated problem is non-convex due to the non-concave objective function with respect to $\boldsymbol{w}$ and $\boldsymbol{\theta}$. In the next section, we adopt a nature-inspired optimizer to solve the problem.

## III. OPTIMIZED WITH HARRIS HAWKS OPTIMIZER

HHO is a population-based and gradient-free optimization technique, which is inspired by the cooperative hunting behaviors of Harris hawks [10]. It can be applied to any optimization problem subject to a proper formulation. The existing literature shows that the HHO provides very promising and competitive results compared with other nature-inspired techniques [10, 11]. The HHO solves optimization problems by mimicking the chasing process of Harris hawks on the prey, such as rabbits, in various scenarios. Mathematically, the chasing activity is modeled into three phases: 1) exploration, 2) exploitation, and 3) transition from exploration to exploitation. In the HHO, the position of Harris hawks, denoted by a vector $X$, represents a candidate solution, and the position of rabbit represents the best solution in each iteration.

*A. Exploration phase*

In this phase, the hawks try to detect a rabbit in the wide solution space based on two strategies: selecting new track positions randomly inside the group's home range or updating their next positions based on the positions of both the other family members and the rabbit. The exploration phase is modeled by

$$X(t+1) = \begin{cases} X_{rand}(t) - r_1 | X_{rand}(t) - 2r_2 X(t) | & q \geq 0.5 \\ X_{rabbit}(t) - X_m(t) - r_3(LB + r_4(UB - LB)) & q < 0.5 \end{cases}, \quad (5)$$

where $X(t)$ and $X(t+1)$ are the position vector of hawks in current and next iteration, respectively, $X_{rand}(t)$ is a randomly selected hawk, $X_m(t)$ is the average position of the current population of hawks, $r_1$, $r_2$, $r_3$, $r_4$ and $q$ are random numbers inside [0, 1] updated every iteration, $LB$ and $UB$ are the lower and upper limits of positions, respectively.

*B. Transition from exploration to exploitation*

The escaping energy of prey affects the hawks' hunting behavior. During the escaping process, the energy $E$ decreases. This fact can be modeled as

$$E = 2E_0(1 - \frac{t}{T}), \quad (6)$$

where $T$ is the maximum number of iterations and $E_0$ is a random number inside the interval [-1, 1] denoting the initial energy in each iteration. When escaping energy $|E| \geq 1$, the HHO works in the exploration phase to explore a rabbit by searching different regions. Otherwise, the optimizer performs a transition between exploration and exploitation, and drives hawks to exploit the neighborhood of the rabbit for a better solution.

*C. Exploitation phase*

In this phase, the hawks try to besiege and kill the rabbit detected in the exploration phase. Meanwhile, the rabbit attempts to escape the roundup. Its escaping probability, denoted by $r$, indicates the chance to successfully escape. The hawks can perform four different chasing strategies according to the rabbit's status represented by escaping energy $E$ and escaping probability $r$.

*1) Soft besiege*

When the rabbit still has enough energy ($|E| \geq 0.5$) and more chance to escape ($r \geq 0.5$), the hawks perform soft besiege strategy to softly encircle the rabbit. This strategy is modeled as

$$X(t+1) = X_{rabbit}(t) - X(t) - E | J \times X_{rabbit}(t) - X(t) |, \quad (7)$$

where $J = 2(1 - r_5)$ represents the random jump strength of the rabbit, and $r_5$ is a random number ranging from 0 to 1. The value of $J$ is updated every iteration.

*2) Hard besiege*

The hard besiege strategy is performed when the rabbit is exhausted ($|E| < 0.5$) but still has a high probability to escape ($r \geq 0.5$). In this situation, the hawks hardly encircle the rabbit and update their positions by

$$X(t+1) = X_{rabbit}(t) - E | X_{rabbit}(t) - X(t) |. \quad (8)$$

*3) Soft besiege with progressive rapid dives*

When $|E| \geq 0.5$ and $r < 0.5$, the rabbit has enough energy to escape. Thus, the hawks adopt a more intelligent strategy to increase the probability of capturing the prey. In this strategy, Levy Flight (LF)-based activities may be adopted as an optimal searching tactic. The positions of hawks are updated by

$$X(t+1) = \begin{cases} Y & \text{if } F(Y) < F(X(t)) \\ Z & \text{if } F(Z) < F(X(t)) \end{cases}, \quad (9)$$

where $Y = X_{rabbit}(t) - E | J \times X_{rabbit}(t) - X(t) |$, $Z = Y + V \times LF(D)$, $LF(\cdot)$ is the levy flight function, $D$ is the dimension of problem, $V$ is a random vector by size $1 \times D$, $F(\cdot)$ is the fitness function derived from the object function of problem.

*4) Hard besiege with progressive rapid dives*

When $|E| < 0.5$ and $r < 0.5$, the rabbit has not enough energy to escape, and the hawks construct a hard besiege and try to decrease the distance between their average location and the rabbit. Thus, the probability of successfully killing the prey is increased. The strategy is modeled as

$$X(t+1) = \begin{cases} Y & \text{if } F(Y) < F(X(t)) \\ Z & \text{if } F(Z) < F(X(t)) \end{cases}, \quad (10)$$

where $Y = X_{rabbit}(t) - E | J \times X_{rabbit}(t) - X_m(t) |$ and $Z = Y + V \times LF(D)$.

*D. Joint beamforming design with HHO*

The HHO algorithm in its original form was designed to optimize the unconstrained problem in real number space. However, the formulated problem (4) is subject to a constraint, and its optimization vector $w$ is a complex vector. Thus, it is necessary to make some transformations on the formulated problem so as to solve it by the HHO.

We first transform the complex vector $w$ into real number vectors and then formulate a new optimization vector $X$, which denotes the search space in the HHO as described above. According to Euler's formula, a complex number can be expressed in the exponential form with a magnitude and an argument. Let $\psi = [\psi_1, \cdots, \psi_M]$ denote magnitude vector with each element being the magnitude of corresponding complex number in $w$. Similarly, let $\varphi = [\varphi_1, \cdots, \varphi_M]$ denote argument vector with each element being the argument of corresponding complex number in $w$. Thus, the optimization vector $X$ in the HHO composes of three parts, that is $X = [\psi, \varphi, \theta]$. It will be converted into $w$ and $\Theta$ when calculating the fitness of the problem.

---

**Algorithm 1** Pseudo-code of HHO algorithm
---
**Inputs**: The population size $Q$ and maximum number of iterations $T$.
**Outputs**: The location of the rabbit and its fitness value.
Initialize the random population $X_i$ ($i = 1, 2, \ldots, Q$), set the iteration index $t$ to 0.
**while** ($t \leq T$) **do**
    Calculate the fitness value for all $X_i$ by (11)
    Set $X_{rabbit}$ as the position of the rabbit (highest fitness value)
    **for**(each $X_i$) **do**
        Update the initial energy $E_0$ and jump strength $J$
        Update escaping energy $E$ by (6)
        **if** ($|E|\geq 1$) **then**
            Update the position vector $X_i(t+1)$ by (5)
        **end if**
        **if** ($|E|< 1$) **then**
            Create a random escaping probability $r$
            **if** ($r \geq 0.5$ and $|E|\geq 0.5$) **then**
                Update the position vector by (7)
            **else if** ($r \geq 0.5$ and $|E|< 0.5$) **then**
                Update the position vector by (8)
            **else if** ($r < 0.5$ and $|E|\geq 0.5$) **then**
                Update the position vector by (9)
            **else if** ($r < 0.5$ and $|E|< 0.5$) **then**
                Update the position vector by (10)
            **end if**
        **end if**
    **end for**
    Set $t = t+1$
**end while**
**return** $X_{rabbit}$ and $F(X_{rabbit})$

---

Furthermore, concerning the constraint, there are several methods to incorporate it in the fitness function, such as penalty method, decoders and feasibility rules [12]. Among these methods, the penalty method is the most common approach in the evolutionary algorithm community, and is also adopted in this work to transform the constrained problem (4) into an unconstrained problem. The main idea of the penalty method is to add a penalty term to the objective function that consists of a penalty parameter multiplied by a measure of violation of the constraints. In particular, the fitness function of problem (4) can be expressed as:

$$F(X) = |(\mathbf{h}_r^H \Theta \mathbf{G} + \mathbf{h}_d^H)\mathbf{w}|^2 + P(X). \quad (11)$$

In (11), $P(\cdot)$ is the penalty term, which can be calculated as

$$P(X) = \begin{cases} 0 & \text{if } \|\mathbf{w}\|^2 \leq P_{AP} \\ -\mu(\|\mathbf{w}\|^2 - P_{AP}) & \text{if } \|\mathbf{w}\|^2 > P_{AP} \end{cases}, \quad (12)$$

where $\mu$ is a penalty factor, which is a positive real number. In this paper, we set $\mu$ to 1. The penalty term is nonzero when the constraint is violated and is zero in other case.

The pseudo code of the HHO algorithm is represented in Algorithm 1.

## IV. SIMULATION RESULTS

We perform numerical simulations to verify the performance of the proposed HHO-based joint beamforming design. The proposed scheme ( Joint AP and IRS beamforming design by the HHO) is compared to the following benchmark schemes: 1) Joint AP and IRS beamforming design by the centralized algorithm proposed in [4, 8], 2) Joint AP and IRS beamforming design by the distributed algorithm proposed in [4, 8], 3) Benchmark scheme without the IRS.

For the sake of performance comparison, the simulation parameters are the same as that in [8]. In the simulation, a two-dimensional plane scenario is considered, and the AP, the IRS and the user are located at (0, 0), (51, 0) and ($d$, 2) in meter (m), respectively. Here $d$ indicates the horizontal distance between the AP and the user. The number of AP's antenna is fixed to 8 ($M$=8), and the reflecting elements at the IRS are placed in a rectangular array with $N = N_x N_y$, where $N_y$ is fixed to 10. Furthermore, it is assumed that the AP-IRS channel is dominated by the Line-of-Sight (LoS) link while both the AP-user and IRS-user channels follow Rayleigh fading with additional 10 dB penetration loss. For all channels, the signal attenuation at a reference distance of 1 meter is set as 30 dB. The antenna gain of AP, user and each reflecting element at the IRS is set to 0, 0, and 5 dBi, respectively. In the simulation, SNR at the user is used as the performance metric. For the HHO-based scheme, $w$ is initialized by $w = \sqrt{P_{AP}} \dfrac{\mathbf{h}_d}{\|\mathbf{h}_d\|}$. Other parameters are set as follows: $\sigma^2$ = -80dBm, $P_{AP}$ = 5dBm, $N_x$ = 5, $\mu$ = 1. All simulations are implemented on a computer with Windows 10 enterprise 64-bit, 32G RAM and 12 logical processors (3.6GHz).

Fig. 1 shows the user SNR of all the schemes in the scenarios with different horizontal distances between the AP and the user, $d$. In this simulation, the results of the HHO-based scheme are obtained by setting the population size $Q$ and the maximum number of iteration $T$ to 80 and 500, respectively. On the one hand, it can be observed from Fig. 1 that the wireless network equipped with the IRS outperforms the system without the IRS in all AP-user distance. The performance improvement becomes very obvious as the AP-user distance increases. The reason is that the user farther away from the AP suffers more signal attenuation, which results in more SNR loss, while the IRS can help the user to increase the received signal power by reflecting signals. The results in Fig. 1 verify that the IRS is able to effectively enhance the performance of the wireless network. On the other hand, it is also observed that the HHO-based scheme performs close to the centralized and distributed scheme. Since both centralized and distributed scheme achieve near optimal SNR, we can conclude that the HHO-based scheme is able to jointly optimize the active and passive beamforming, and obtain a near optimal solution. Furthermore, as the HHO is a well-established optimization technique, the HHO-based scheme is easier than the other two schemes in the design and implementation phase.

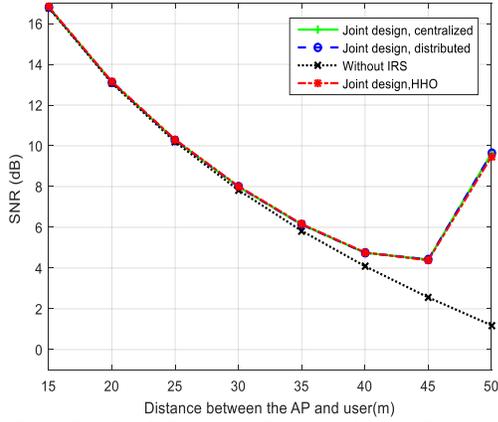

Fig. 1. Receive SNR versus AP-user horizontal distance, $d$

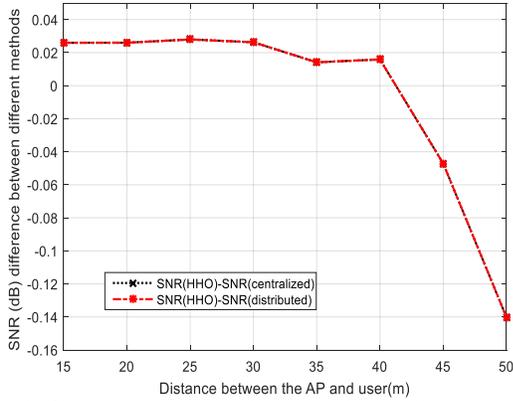

Fig. 2. Performance difference between different algorithms ($Q$=80, $T$=500)

The performance differences between the centralized, the distributed and the HHO-based schemes are more clearly shown in Fig. 2, where each data point is the difference between the user SNR obtained by the HHO-based scheme minus the user SNR obtained by other schemes. A positive number indicates that the HHO-based scheme obtains better solutions, whereas a negative number means other schemes have better results. It can be observed in Fig. 2 that the HHO-based scheme achieves a slightly better solution than centralized and distributed schemes in most AP-user distance ($d\leq40$) with current settings ($Q$=80,

$T$=500). When the user is nearer to the AP, the received signal at the user is dominated by the AP-user direct link. Thus, the initialization of $w$ based on AP-user channel information helps the HHO to find better solutions quickly. As the user is near to the IRS, both AP-IRS and IRS-user link is dominant. The HHO algorithm requires more iterations or a larger population size to increase the optimality of results.

Figs 3 and 4 are the convergence curve of the HHO for $d$=25 and 50 with $T$=1500, respectively, which show the fitness of the best solution found during iteration. The curves in the two figures reveal an accelerated convergence trend, and most of the convergence is done before the first 500 iterations. The quality of the solution can continue to be slowly improved by more iterations and a larger population size. This can be verified by comparing Fig. 2 and Fig. 5. In the two figures, the results are obtained with $T$=500, $Q$=80 and $T$=1500, $Q$=200, respectively. As shown in Fig. 5, the HHO-based scheme outperforms other schemes in all AP-user distance. With more iterations and a larger population size, the HHO can obtain better results. However, more iterations and a larger population size consumes more time. As shown in Fig. 6, the time consumed in the HHO-based scheme is linearly related to the number of iterations and population size. Therefore, it is cost-effective to obtain a near optimal solution by empirically setting a reasonable number of iterations and population size.

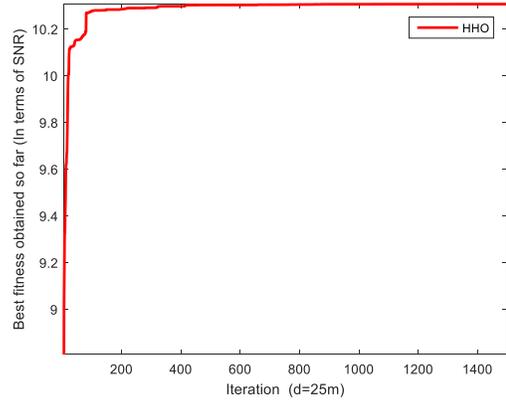

Fig. 3. Convergence curve of the proposed HHO-based scheme ($d$=25, $Q$=80, $T$=1500)

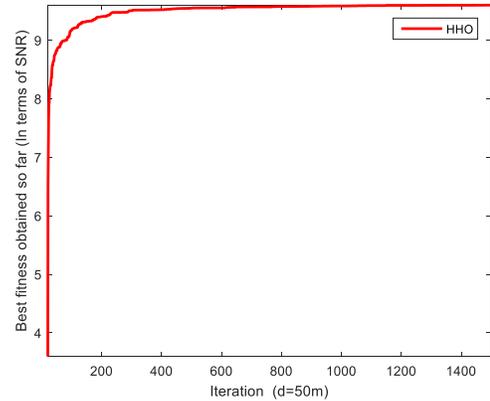

Fig. 4. Convergence curve of the proposed HHO-based scheme ($d$=50, $Q$=80, $T$=1500)

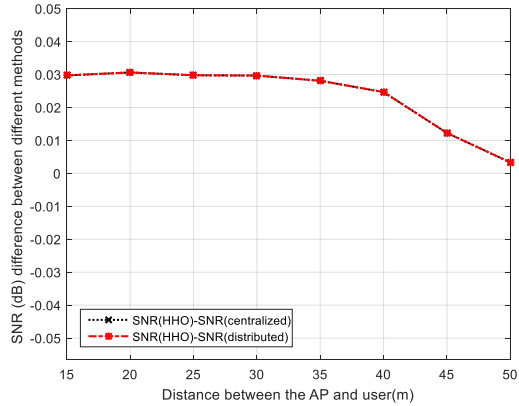

Fig. 5. Performance difference between different schemes($Q$=200, $T$=1500)

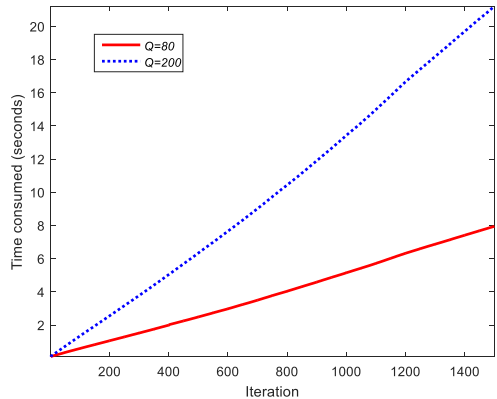

Fig. 6. Time consumed with different iterations and population size

## V. CONCLUSION

In this paper, we use a nature-inspired optimization technique, Harris hawks optimizer, to jointly optimize the transmit beamforming at the AP and the passive reflect beamforming at the IRS to maximize the user's received signal power in an IRS-aided MISO wireless network. After incorporating the constraint into the fitness function, the formulated non-convex optimization problem is optimized by the HHO algorithm. The simulation results show that the HHO-based scheme could obtain similar or even better results compared with other algorithms.

## ACKNOWLEDGMENT

This research was supported by 1) Nanyang Technological University (NTU) Startup Grant, 2) Alibaba-NTU Singapore Joint Research Institute (JRI), 3) Singapore Ministry of Education Academic Research Fund Tier 1 RG128/18, Tier 1 RG115/19, Tier 1 RT07/19, Tier 1 RT01/19, and Tier 2 MOE2019-T2-1-176, 4) NTU-WASP Joint Project, 5) Singapore National Research Foundation (NRF) under its Strategic Capability Research Centres Funding Initiative: Strategic Centre for Research in Privacy-Preserving Technologies & Systems (SCRIPTS), 6) Energy Research Institute @NTU (ERIAN), 7) Singapore NRF National Satellite of Excellence, Design Science and Technology for Secure Critical Infrastructure NSoE DeST-SCI2019-0012, 8) AI Singapore (AISG) 100 Experiments (100E) programme, 9) NTU Project for Large Vertical Take-Off & Landing (VTOL) Research Platform. 10) Shandong Province Higher Educational Science and Technology Program (J16LN21), 11) Nantong Science and Technology Project (MS12018082), and 12) Study Abroad Program by the Government of Shandong Province (201801014).

## REFERENCES

[1] Q. Wu, G. Y. Li, W. Chen, D. W. K. Ng, and R. Schober, "An overview of sustainable green 5G networks," *IEEE Wireless Communications,* vol. 24, pp. 72-80, Aug. 2017.
[2] S. Zhang, Q. Wu, S. Xu, and G. Y. Li, "Fundamental green tradeoffs: Progresses, challenges, and impacts on 5G networks," *IEEE Communications Surveys & Tutorials,* vol. 19, pp. 33-56, July 2016.
[3] M. Di Renzo, M. Debbah, D.-T. Phan-Huy, A. Zappone, M.-S. Alouini, C. Yuen, V. Sciancalepore, G. C. Alexandropoulos, J. Hoydis, and H. Gacanin, "Smart radio environments empowered by reconfigurable AI meta-surfaces: An idea whose time has come," *EURASIP Journal on Wireless Communications and Networking,* vol. 2019, pp. 1-20, May. 2019.
[4] Q. Wu and R. Zhang, "Intelligent reflecting surface enhanced wireless network via joint active and passive beamforming," *IEEE Transactions on Wireless Communications,* vol. 18, pp. 5394-5409, Nov. 2019.
[5] P. Wang, J. Fang, X. Yuan, Z. Chen, H. Duan, and H. Li. Intelligent reflecting surface-assisted millimeter wave communications: Joint active and passive precoding design [Online]. Available: https://arxiv.org/abs/1908.10734
[6] S. Zhang and R. Zhang, "Capacity characterization for intelligent reflecting surface aided MIMO communication," *IEEE Journal on Selected Areas in Communications,* 2020.
[7] S. Gong, X. Lu, D. T. Hoang, D. Niyato, L. Shu, D. I. Kim, and Y.-C. Liang, "Towards Smart Wireless Communications via Intelligent Reflecting Surfaces: A Contemporary Survey," *IEEE Communications Surveys & Tutorials,* 2020.
[8] Q. Wu and R. Zhang, "Intelligent reflecting surface enhanced wireless network: Joint active and passive beamforming design," in *2018 IEEE Global Communications Conference (GLOBECOM)*, Abu Dhabi, United Arab Emirates, 2018, pp. 1-6.
[9] M. Cui, G. Zhang, and R. Zhang, "Secure wireless communication via intelligent reflecting surface," *IEEE Wireless Communications Letters,* vol. 8, pp. 1410-1414, Oct. 2019.
[10] A. A. Heidari, S. Mirjalili, H. Faris, I. Aljarah, M. Mafarja, and H. Chen, "Harris hawks optimization: Algorithm and applications," *Future Generation Computer Systems,* vol. 97, pp. 849-872, Aug. 2019.
[11] Q.-V. Pham, T. Huynh-The, M. Alazab, J. Zhao, and W.-J. Hwang, "Sum-Rate Maximization for UAV-assisted Visible Light Communications using NOMA: Swarm Intelligence meets Machine Learning," *IEEE Internet of Things Journal,* 2020.
[12] A. R. Jordehi, "A review on constraint handling strategies in particle swarm optimisation," *Neural Computing and Applications,* vol. 26, pp. 1265-1275, Jan. 2015.